\documentclass[10pt, conference]{IEEEtran}
\IEEEoverridecommandlockouts

\usepackage[utf8]{inputenc}
\usepackage[english]{babel}
\usepackage{subcaption}
\usepackage{graphicx}
\usepackage{xcolor}
\usepackage{cite}
\usepackage{amsmath}
\usepackage{multirow}
\usepackage{bm}
\usepackage{array}
\usepackage{url}
\usepackage{psfrag}
\usepackage{mathtools}
\usepackage{amssymb}
\usepackage{algpseudocode}
\usepackage[normalem]{ulem}
\usepackage{soul}
\usepackage[linesnumbered,ruled]{algorithm2e}
\usepackage{comment}
\usepackage{siunitx}
\usepackage{booktabs}
\usepackage{lipsum}
\usepackage{mathtools}
\usepackage{cuted}
\usepackage{floatrow}
\usepackage[acronym]{glossaries}
\usepackage{bbm}

\include{abberviaiton.tex}

\newfloatcommand{capbtabbox}{table}[][\FBwidth]

\newcommand{\vect}{\boldsymbol}

\newcommand{\indict}[1]{\mathbbm{1}(#1)}
\newcommand{\rActual}{\boldsymbol{\gamma}}

\newcommand{\bActual}{\boldsymbol{\rho}}
\newcommand{\bPredict}{\hat{\bActual}}

\def\BibTeX{{\rm B\kern-.05em{\sc i\kern-.025em b}\kern-.08em
    T\kern-.1667em\lower.7ex\hbox{E}\kern-.125emX}}
\begin{document}


\newacronym{ofdm}{OFDM}{orthogonal frequency-division multiplexing}
\newacronym{rx}{Rx}{receiver}
\newacronym{tx}{Tx}{transmitter}
\newacronym{ai}{AI}{artificial intelligence}
\newacronym{ula}{ULA}{uniform linear array}
\newacronym{bs}{BS}{base station}
\newacronym{lidar}{LiDAR}{light detection and ranging}
\newacronym{fov}{FoV}{field-of-view}
\newacronym{rf}{RF}{radio frequency}
\newacronym{dbscan}{DBSCAN}{density-based spatial clustering of applications with noise}
\newacronym{mmwave}{mmWave}{millimeter-wave}
\newacronym{qos}{QoS}{quality of service}
\newacronym{iot}{IoT}{internet of things}
\newacronym{lstm}{LSTM}{long short-term memory}
\newacronym{smae}{sMAE}{smooth mean absolute error}
\newacronym{src}{SRC}{static clutter removal}
\newacronym{rssi}{RSSI}{received signal strength indicator}
\newacronym{cdf}{CDF}{cumulative distribution function}
\newacronym{dnn}{DNN}{deep neural network}
\newacronym{nn}{NN}{neural network}
\newacronym{cnn}{CNN}{convolutional neural network}
\newacronym{rnn}{RNN}{recurrent neural network}
\newacronym{relu}{ReLU}{rectified linear unit}
\newacronym{bce}{BCE}{binary cross-entropy}
\newacronym{dl}{DL}{deep learning}
\newacronym{ssl}{SSL}{self-supervised learning}

\title{Zero-Shot Generalization for Blockage Localization in mmWave Communication}
\author{
\IEEEauthorblockN{
Rafaela~Scaciota\IEEEauthorrefmark{1}\IEEEauthorrefmark{2},~\IEEEmembership{Member,~IEEE,} Malith~Gallage\IEEEauthorrefmark{1},~\IEEEmembership{Student Member,~IEEE,}
Sumudu~Samarakoon\IEEEauthorrefmark{1}\IEEEauthorrefmark{2},~\IEEEmembership{Member,~IEEE},
\\and
Mehdi Bennis\IEEEauthorrefmark{1},~\IEEEmembership{Fellow,~IEEE}
}
\IEEEauthorblockA{
	\small%
	\IEEEauthorrefmark{1}%
	Centre for Wireless Communication, University of Oulu, Finland \\
	\IEEEauthorrefmark{2}%
 Infotech Oulu, University of Oulu, Finland \\
 email: \{rafaela.scaciotatimoesdasilva, malith.gallage, sumudu.samarakoon, mehdi.benni\}@oulu.fi
}

}
\vspace{-20pt}

\maketitle

\begin{abstract}
This paper introduces a novel method for predicting blockages in \gls{mmwave} communication systems towards enabling reliable connectivity. It employs a self-supervised learning approach to label \gls{rf} data with the locations of blockage-causing objects extracted from \gls{lidar} data, which is then used to train a deep learning model that predicts object's location only using RF data. Then, the predicted location is utilized to predict blockages, enabling adaptability without retraining when transmitter-receiver positions change. Evaluations demonstrate up to $74 \%$ accuracy in predicting blockage locations in dynamic environments, showcasing the robustness of the proposed solution.

\end{abstract}

\glsresetall
\section{Introduction}

\Gls{mmwave} communication has emerged as a promising technology for next-generation wireless networks due to its ability to offer high data rates with low latency. However, \gls{mmwave} signals are susceptible to blockages caused by objects in the environment, which can lead to signal attenuation and link failures~\cite{Rappaport.17}. Therefore, predicting and mitigating blockages in \gls{mmwave} communication systems have become critical to ensure reliable and uninterrupted connectivity. Anticipating the blockages in advance allows proactive measures to maintain signal integrity and enhance \gls{qos} through mechanisms such as handover, beamforming, caching, and others\cite{Yang.20}.

The recent advancements in wireless communication technologies, especially in the \gls{mmwave} bands, have significant research interest in blockage prediction. 
Building on the foundational propagation models and performance metrics for $5$G \gls{mmwave} bands established in~\cite{Sun.18}, researchers have explored innovative applications of \gls{mmwave} technology beyond traditional communication. For example, \cite{Huang.22} showcased the use of \gls{mmwave} radar for indoor object detection and tracking, while \cite{Koda.20} introduced multimodal split learning techniques to improve \gls{mmwave} received power prediction in resource-constrained settings. Expanding further, \cite{Jiang.23} utilized \gls{lidar} to predict mobile blockages and optimize beam computation, and \cite{Alloulah.22} proposed advanced learning techniques such as distributed heteromodal split learning and self-supervised radio-visual representation learning to refine prediction and sensing capabilities. Together, these advancements underscore the potential of integrating diverse sensing modalities to enhance the robustness and functionality of \gls{mmwave} communication systems.

The application of \gls{mmwave} and multisensor technologies goes beyond traditional communication paradigms. In~\cite{Yang.23}, a self-supervised geometric learning framework for automatic Wi-Fi human sensing is introduced for \gls{iot} applications. Furthermore, the authors in~\cite{Saeed.21} explored self-supervised federated learning approaches for multisensor representations, offering scalable and privacy-preserving embedded intelligence solutions in various sensor networks. In~\cite{wu.23}, the authors use \gls{lidar} sensory data to detect whether an object will block the \gls{mmwave} communication link between \gls{tx} and \gls{rx}. The authors in~\cite{Wu.22} use a deep learning architecture proposed to learn the pre-blockage signal in the \gls{rf} and \gls{lidar} data and successfully predict blockages proactively. One drawback of the previous studies is that even a slight change in the position of the \gls{tx} or \gls{rx} requires the model to be re-trained and thus, developing robust and adaptive blockage prediction techniques are paramount for \gls{mmwave} communication system designs.

 The main contribution of this paper is to propose a \textbf{novel blockage prediction solution for \gls{mmwave} communication systems using stressing cross-\gls{ssl}} that can be easily adapted for certain modifications in the environments. Towards this, we first employ a \gls{ssl} approach to label the \gls{rf} data with the locations of objects causing the blockages that are extracted from \gls{lidar} data. Using them, we train a \gls{lstm} neural network to predict future object locations. Then, a geometric analysis is carried out to evaluate whether a blockage will occur at the predicted location at any given time. The proposed solution can be used to predict the blockages at newly deployed \gls{tx}-\gls{rx} links in the vicinity, by simply reconfiguring the geometric analysis without a need of retraining the \gls{dl} model. Finally, we present a real-world evaluation of a point-to-point communication scenario using the deepSense6G dataset \cite{DeepSense}. This paper is organized as follows. 
 Section~\ref{sec:system} presents the system model and the problem formulation. The methodology is discussed in Section~\ref{sec:methodology}.
 Section~\ref{sec:results} lays out the evaluation of the proposed method while Section~\ref{sec:conclusion} draws the conclusions.
 
\section{System Model \& Problem Formulation}\label{sec:system}

\begin{figure}[!t]
    \centering
    \includegraphics[width=1\linewidth]{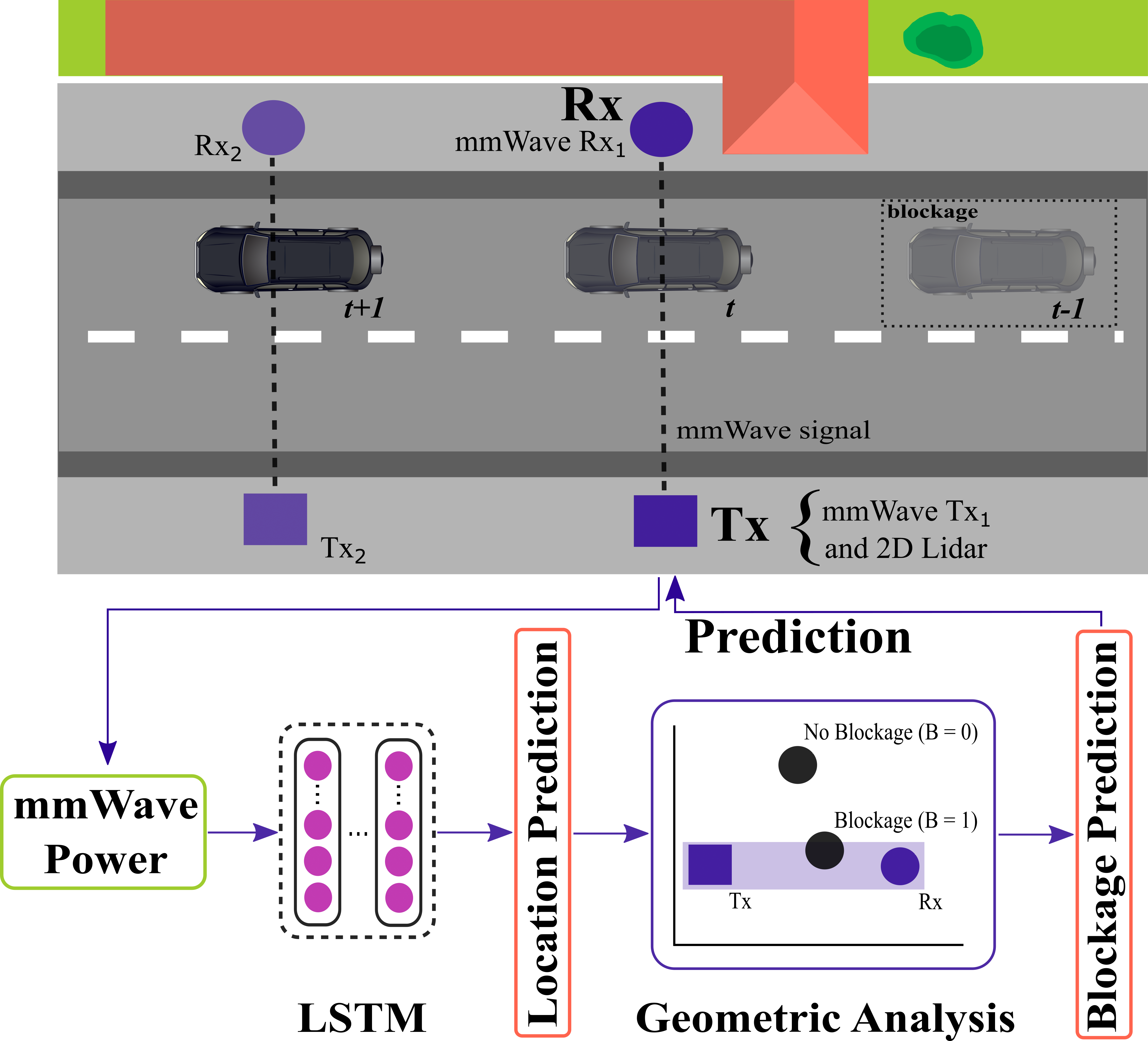}
    \caption{Top: Overall system model where a \gls{mmwave} base station leverages both \gls{lidar} and in-band wireless sensing to predict the object location coordinates. Bottom: The flow diagram highlighting the key steps of the proposed solution.}
    \label{fig:systemModel}
\end{figure}

We consider a point-to-point \gls{mmwave} communication system consisting of a \gls{tx} $M_A$-element \gls{ula} antenna and a static \gls{rx} with a single antenna. The \gls{tx} is equipped with a $2$D \gls{lidar} that provides situational awareness about the environment and any moving objects, as shown in Fig.~\ref{fig:systemModel}. The system uses a predetermined beam steering codebook consisting of $M$ beams, denoted as $\mathcal{F} = \{ \boldsymbol{f}_m\}^M_{m=1}$. 
Each beam steering vector $\boldsymbol{F}_m = \boldsymbol{f}_m(\Theta_m )$
is designed to direct signals in a specific direction given by $\Theta_m = \Theta_{\text{offset}} + (F_{\text{off-view}}/M)$, where $\Theta_{\text{offset}}$ is the signal direction offset and, $F_{\text{off-view}}$ denotes the \gls{fov} of the wireless beamforming system. 

The signal transmission model is \gls{ofdm} with $K$ subcarriers. We denote $\boldsymbol{h}_{k,t} \in \mathbb{C}^{M \times 1}$ as the downlink channel from \gls{tx} to \gls{rx} on the $k$\textsuperscript{th} subcarrier at time $t$, where $t \in \mathbb{Z}$. 
At time $t$, if the \gls{tx} adopts the beam steering vector $\boldsymbol{F}_m$ for downlink transmission, the received signal at subcarrier $k$ is represented as
\begin{equation}
    r_{k,m,t} = \boldsymbol{h}_{k,t}^T \boldsymbol{F}_m s_{k,t} + n_{k,t},
\end{equation}
where $s_{k,t}$ represents the transmitted symbol at the $k$\textsuperscript{th} subcarrier 
with $\mathbb{E}[s_{k,t}]^2 = 1$, and $n_{k,t} \thicksim \mathcal{CN}(\mathbf{0}, \sigma_n^2)$ is complex Gaussian noise with zero mean and the covariance of $\sigma_n^2$.
In this view, the \gls{rssi} vector of the $M$ beams at time $t$ can be defined as $\boldsymbol{r}_t = [ |\boldsymbol{r}_{1,t}|^2,...,|\boldsymbol{r}_{M,t}|^2 ]^T \in\mathcal{S}$, where $\mathcal{S}$ is the \gls{rssi} observation space, and $|\boldsymbol{r}_{m,t}|^2 = \sum_{k=1}^{K} |r_{k,m,t}|^2$ represents the total power received by the $m$\textsuperscript{th} beamforming vector over $K$ subcarriers. We assume that a collection $\vect{S}_t = [\vect{r}_{t-\tau}]_{\tau\in\mathcal{T}}\in \mathcal{S}^{T_0}$ of past $T_0$ \gls{rssi} observations are drawn from a bounded space where $\mathcal{T} = \{0, 1, \ldots, T_0\}$.

Simultaneously, the \gls{lidar} sensor in \gls{tx} provides a snapshot of the surrounding environment at each time step $t$ through a data matrix  $\boldsymbol{\ell}_{t} \in \mathcal{L}$ where $\mathcal{L}$ defines the \gls{lidar} observation space. Each row represents a single \gls{lidar} point characterized by the angle related to the sensor and the depth value measured at the corresponding angle. Similar to \gls{rssi}, we define a collection of $T_0$ past \gls{lidar} observations at time $t$ as $\vect{L}_t=[\boldsymbol{\ell}_{t-\tau}]_{\tau \in \mathcal{T}} \in \mathcal{L}^{T_0}$.

In a \gls{mmwave} communication setting, a blockage is defined by an event where the resultant signal strength at the \gls{rx} over $K$ subcarriers is below a predefined threshold $r_0$. In this view, we formally define an indication of a blockage at a given time $t$ as $b_t = \indict{|\boldsymbol{r}_t| < r_0}$ where $\indict{\cdot}$ is the indicator function. The conventional data-driven designs of blockage predictions over a future horizon of length $N$ rely on the past sensory information by learning a mapping function $U_{\vect{\theta}}:\mathcal{S}^{T_0}\times\mathcal{L}^{T_0}\to\{0,1\}^N$ with parameters $\vect{\theta}$. Therein, the learning objective can be formalized as follows:
\begin{equation}
    \min_{\vect{\theta}} \lim_{T\to\infty} \frac{1}{T} \sum_{t=0}^T \frac{\| U_{\theta}(\vect{S}_t, \vect{L}_t) - \vect{\rho}_t \|^2}{N},
    \label{eq:out_probl}
\end{equation}
where $\bPredict_t = [\hat{b}_{t+1}, \dots, \hat{b}_{t+N}]$ and $\bActual_t = [{b}_{t+1}, \dots, {b}_{t+N}]$ are the predicted and actual future blockages, respectively.

In contrast to exploiting the correlation between \gls{rssi}, \gls{lidar} data and blockage, we seek a method to utilize the dependencies of sensory data and object locations to predict future locations of the object and derive future blockages based on the geometry of \gls{tx}-\gls{rx} link and object locations. It is worth noting that the object location is unavailable and, thus, needs to be extracted from a mapping $V_{\vect{\phi}}: \mathcal{L} \to \mathcal{X}$ with parameter $\vect{\phi}$ in a self-supervised manner. Here, $\mathcal{X}$ represents the space of the object locations. Then, the future locations for $N$ time horizon can be predicted via learning a $\vect{\psi}$-parametric function $W_{\vect{\psi}} : \mathcal{X} \times \mathcal{S}^{T_0} \to \mathcal{X}^N$. Learning $W_{\vect{\psi}}$ is carried out with the following objective:
\begin{equation}
    \min_{\vect{\phi},\vect{\psi}} \lim_{T\to\infty} \frac{1}{T} \sum_{t=0}^T \frac{  \Gamma( W_{\psi}(\vect{S}_t, V_{\phi}(\vect{\ell}_t) ), \vect{\gamma}_t) }{N},
    \label{eq:pred_probl}
\end{equation}
where $\rActual_t = [ V_{\phi}(\vect{\ell}_{t + \tau}) ]_{\tau\in\{1, \dots, N\}}$ is the future locations of the object, and $\Gamma$ is the loss function.

\section{Deep Learning Solution for Blockage Prediction}\label{sec:methodology}

The problem described in \eqref{eq:pred_probl} can be cast as a deep learning problem due to the inherent capability to model complex, non-linear relationships in data. In this context, the goal is to understand the relationships between object locations and blockage events, making it well-suited for analyzing blockage dynamics and predicting future blockages based on the past observations.

To predict future blockages, our deep learning approach comprises two stages:
(i) creating a labeled dataset of \gls{rssi} and location data through \gls{ssl} and
(ii) implementing the deep learning solution to predict blockages based on object location. The experimental setup used to evaluate the proposed deep learning solution, including the \gls{mmwave} communication system and the associated testbed, was developed by the deepSense6G project~\cite{DeepSense}.

\subsection{Self-supervised approach for dataset generation}\label{sub:data_cre}

To improve data quality and isolate relevant information, we applied the \gls{src} noise filtering algorithm to the raw \gls{lidar} data as proposed by the authors in~\cite{shunyao22}. We filter out potential noise by removing the \gls{lidar} points within a defined proximity to \gls{tx} and isolate the road section by removing all other surrounding data. This ensures that we filter out the \gls{lidar} points that do not belong to objects on the road. 

Then, we employ an unsupervised learning approach to label \gls{rf} data, allowing the grouping of \gls{lidar} data points into distinct categories based on their similarities to estimate the location of the object as $\rActual_t =  V_{\phi}(\vect{\ell}_t)$. We apply the \gls{dbscan} unsupervised learning framework, which is known for its ability to cluster data points densely packed in high-density regions~\cite{ester.96}. The dataset is generated with the collection of $T_0$ such observations, which is denoted by $\mathcal{D} = \{(\boldsymbol{S}_t,  \rActual_t)\}_{t\in\{1,\dots,T_0\}}$.

\subsection{Blockage Prediction based on Object Location}\label{sub:predict}

First, we predict the future locations of the object by learning the mapping $W_{\vect{\psi}}$ using a deep learning model. Among the most efficient types of deep learning for activity prediction, \gls{lstm} stands out due to its unique ability to remember single events over long and often unknown periods. Here, the \gls{lstm} output is a scalar representing the object location. During the training period, given the fixed length of the observation window $T_i$, let $\boldsymbol{I}_t=[(\boldsymbol{S}_t, \rActual_t), ..., (\boldsymbol{S}_{t - T_i + 1}, \rActual_{t - T_i + 1})]$, denote an input data sample. As for the loss function, we use \gls{smae}, also known as Huber Loss, given by
\begin{equation}\label{eq:loss}
    \Gamma(\vect{Z}_t)= \begin{cases}
    \frac{1}{2}\vect{Z}_t^2, & \text{if~} |\vect{Z}_t| \leq\delta,\\
        \delta (|\vect{Z}_t| - \frac{1}{2}\delta) , & \text{otherwise},
        \end{cases}
\end{equation}
 where $\vect{Z}_t =W_{\psi}(\boldsymbol{I}_t) - \vect{\gamma}_t$, and $\delta \in \mathbb{R}^{+}$ is the parameter that determines the threshold for switching between the quadratic and linear components of the loss function. This makes the function less sensitive to outliers, but it also leads to penalization of minor errors within the data sample. Adam optimizer obtains the optimal choices for $\phi$ and $\psi$. Finally, to predict the blockage, we apply a geometric analysis that incorporates the vehicle location as
\begin{equation}\label{eq:pred_block}
    B_t = \begin{cases}
    1, & \text{if~} 0\leq v\leq 1 \text{~and~} 0\leq u\leq 1,\\
        0, & \text{otherwise},
        \end{cases}
\end{equation}
where $v = (\hat{y}_\text{B} - y_\text{T})/(y_\text{R} - y_\text{T})$ and $u = 1/2 + (1/\omega) (\hat{x}_\text{B} - y_\text{T} - t)$. Here, the predicted object location is $\hat{\gamma}_t=(\hat{x}_\text{B}, \hat{y}_\text{B})$, the object width is $\omega$, and the locations of \gls{tx} and \gls{rx} are $(x_\text{T}, y_\text{T})$ and $(x_\text{R}, y_\text{R})$, respectively.

\begin{figure}{}
      \centering
	     \begin{subfigure}{0.48\linewidth}
		 \includegraphics[width=1.1\textwidth]{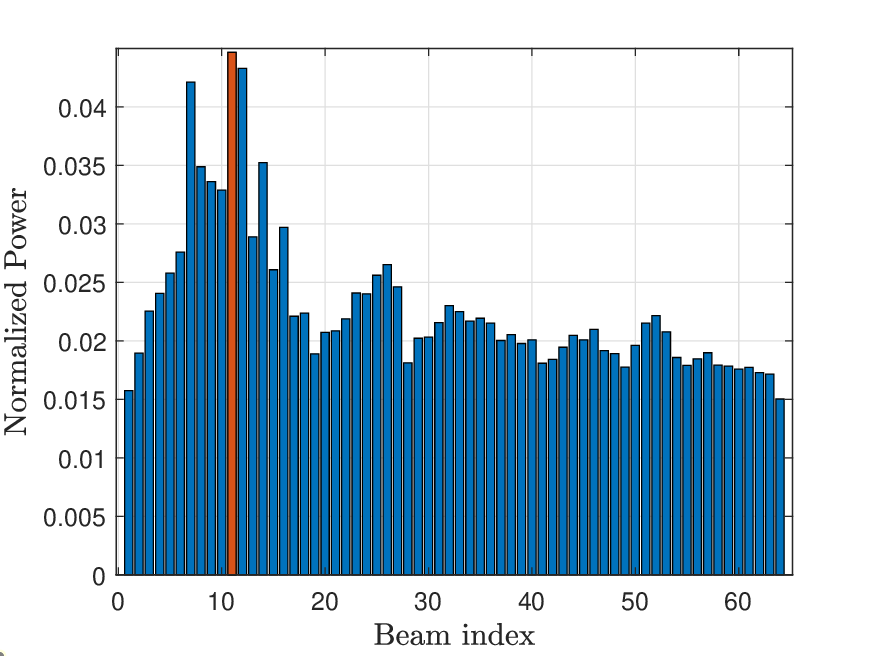}
            \caption{\gls{mmwave} Beam Power}
            \label{fig:mmwave}
	      \end{subfigure}
            \begin{subfigure}{0.48\linewidth}
	          \includegraphics[width=1.2\textwidth]{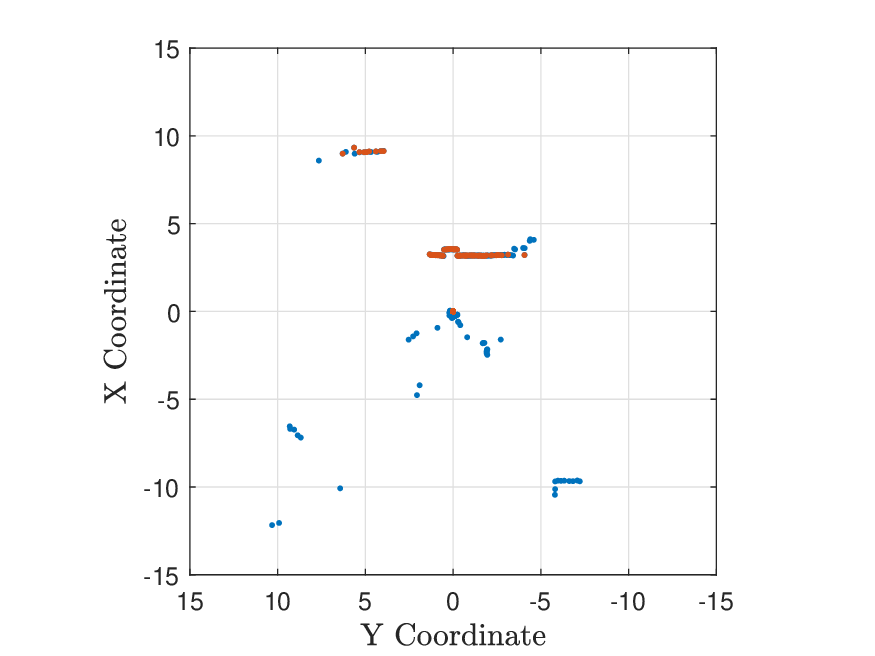}
                \caption{2D \gls{lidar}}
                \label{fig:lidar}
	       \end{subfigure}
	\caption{DeepSense 6G dataset visualization. a) Blue bar: average \gls{rssi} over the $16 \times 16$ beams. Red bar: the \gls{rssi} of an arbitrarily selected beam. b) Blue points: raw \gls{lidar} data. Red points: isolated relevant \gls{lidar} data.}
\label{fig:data}
\end{figure}

\section{Performance Evaluation}\label{sec:results}

\begin{table}[]
\begin{tabular}{|l|ccc|}
\hline
\multicolumn{1}{|c|}{\multirow{2}{*}{\textbf{Parameter}}} & \multicolumn{1}{c|}{\textbf{\begin{tabular}[c]{@{}c@{}}RF\\ 
Blockage\end{tabular}}} & \multicolumn{1}{c|}{\textbf{\begin{tabular}[c]{@{}c@{}}Lidar + RF\\ 
Blockage\end{tabular}}} & \textbf{\begin{tabular}[c]{@{}c@{}}RF\\
Localization\end{tabular}} \\ \cline{2-4} 
\multicolumn{1}{|c|}{}                                    & \multicolumn{3}{c|}{\textbf{Value}}  \\ \hline
Input size                                                & \multicolumn{1}{c|}{$64$}   & \multicolumn{1}{c|}{$64$}    & $64$   \\ \hline
Hidden Size                                               & \multicolumn{1}{c|}{$16$}   & \multicolumn{1}{c|}{$16$}    & $32$   \\ \hline
\gls{lstm} Layers                                         & \multicolumn{1}{c|}{$4$}    & \multicolumn{1}{c|}{$2$}     & $1$     \\ \hline
Learning rate                                             & \multicolumn{1}{c|}{$1e{-3}$}  & \multicolumn{1}{c|}{$1e{-3}$}    & $1e{-3} $     \\ \hline
Batch size                                                & \multicolumn{1}{c|}{$8$}    & \multicolumn{1}{c|}{$8$}     & $8$    \\ \hline
Loss function                                             & \multicolumn{1}{c|}{BCE Loss}  & \multicolumn{1}{c|}{BCE Loss}   & Huber Loss  \\ \hline
Episodes                                                  &\multicolumn{3}{c|}{$10$} \\ \hline
Iterations                                                &\multicolumn{3}{c|}{$100$} \\ \hline
\end{tabular}
\caption{\Gls{dnn} model parameters.}%
  \label{tab:lstm_parameters}
\end{table}

For the evaluations, we rely on the data from the deepSense6G project~\cite{DeepSense}, a large-scale data set that captures real-world \gls{lidar}, communication, and positional data for research. More specifically, we adopt the scenarios $24-29$ in deepSense6G testbed $3$ to create the dataset.
The testbed $3$ features a \gls{tx} and a \gls{rx}, each equipped with $60\,$GHz \gls{mmwave} phased arrays consisting of $16$-element \gls{ula}. 
The communication uses \gls{ofdm} modulation with a bandwidth of $20$ MHz and $64$ subcarriers. The \gls{lidar} is mounted on the \gls{tx}. The scanning range of \gls{lidar} is $16$ meters and the motor spin frequency is $10$Hz. At each capture, \gls{lidar} creates a $360^\circ$ point cloud. 

A snapshot of data used in this work is presented in Fig.~\ref{fig:data}. 
Fig.~\ref{fig:mmwave} shows the average \gls{rssi} over the $16 \times 16$ beams as the blue bar and the red bar is the \gls{rssi} of an arbitrarily selected beam. The bird's eye view constructed using the $2$D \gls{lidar} data is presented in Fig.~\ref{fig:lidar}. Here, the raw data from \gls{lidar} is shown as blue points, and the isolated relevant \gls{lidar} data is shown in red. 

To structure the data prepared for vehicle location prediction, we used a sliding window method with an observation window size of $T_0 = 8$.  The sequenced data were then used to train a \gls{lstm} network to predict the location of the vehicle centroids for the following $N = 5$ time windows. We used a distance threshold of $\epsilon = 2$ to determine the neighbor relationships between points to cluster data points. Furthermore, we established a minimum sample size of $4$ points required to form a cluster. 

The \gls{lstm} is configured with a hidden layer that determines the dimensionality of its hidden state. The output of the \gls{lstm} layers is arranged so that the batch size is the first dimension, which is then processed through two fully connected (dense) layers. The first dense layer reduces the dimensionality from $64$ to $20$ neurons and applies a \gls{relu} activation function to introduce non-linearity. The output of this layer is passed to the second dense layer, which further reduces the output to $2N$ neurons. A final \gls{relu} activation function is applied to produce the final output. We denote the proposed framework as \textbf{RF-Localization} to explicitly indicate that the proposed framework utilizes received power values for the localization prediction, then evaluate the blockage  applying~\eqref{eq:loss}.

The proposed prediction framework is evaluated against two baseline models that rely exclusively on either \gls{rssi} sequences and/or \gls{lidar} sequences. In the first baseline model, referred to as \textbf{RF-Blockage}, only the \gls{rssi} values are fed into the \gls{lstm} layer. This model uses the same architecture as the \textbf{RF-Localization} model; however, the labels are the blockage states and evaluated with a \gls{bce} loss. Then, the blockage event needs to be extracted from a mapping as $H_{\vect{\phi}}: \mathcal{L} \to \mathcal{B}$ with parameter $\vect{\phi}$ in a self-supervised manner. Here, $\mathcal{B}$ represents the space of the blockage events. As a result, the model learns a mapping function, $U_{\vect{\theta}}:\mathcal{S}^{T_0}\times H_{\phi}(\vect{\ell}_t)\to\{0,1\}^N$, where the output represents the predicted blockage states 

In the second baseline, termed as \textbf{RF+LiDAR-Blockage}, we also predicted blockage states using a multimodal framework that utilizes both \gls{lidar} and \gls{rssi} values for the prediction by learning a mapping function $U_{\vect{\theta}}:\mathcal{S}^{T_0}\times\mathcal{L}^{T_0}\to\{0,1\}^N$. The multimodal framework integrates two components: a \gls{cnn} and a \gls{lstm}. The \gls{cnn} processes \gls{lidar} data, leveraging convolutional layers to extract spatial features and patterns. Meanwhile, the \gls{lstm} analyzes sequential \gls{rf} data, capturing temporal dependencies. The outputs from these two networks are concatenated into a unified feature vector, which merges both spatial and temporal information. This combined vector is then fed into a fully connected layer that transforms it into a final output of size $2N$. To produce the final predictions, a sigmoid activation function is applied to the output, generating probability values that can be interpreted as the model’s predictions. It is worth highlighting that \textbf{RF+Lidar-Blockage} model is evaluated with a \gls{bce} loss.

\begin{figure}[t]
    \centering
    \includegraphics[width=1\linewidth]{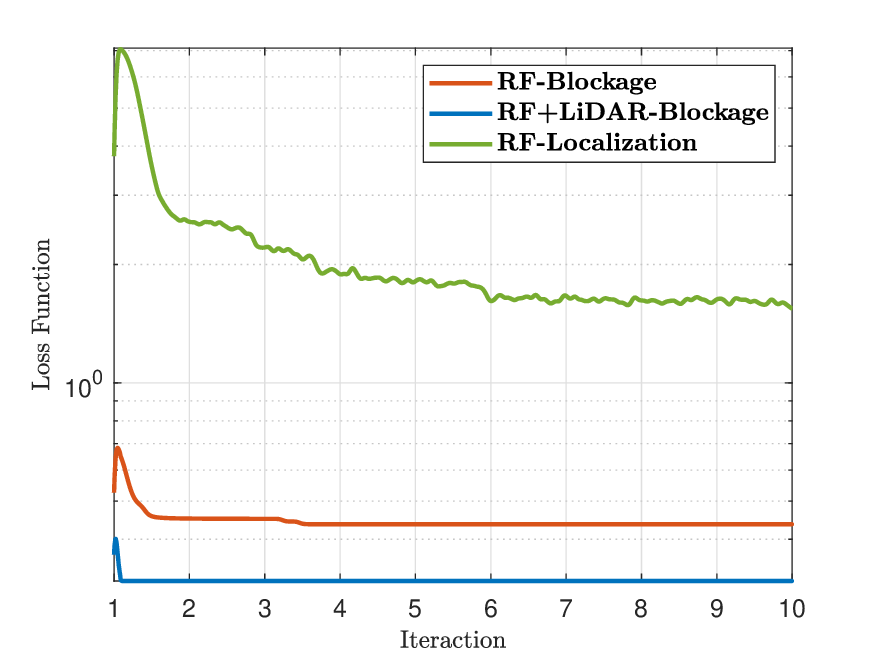}
    \caption{Training loss \gls{lstm} model using \textbf{RF-Blockage} method, \textbf{RF+LiDAR-Blockage} method and the proposed method, \textbf{RF-Localization}.}
    \label{fig:train_loss}
\end{figure}

The model parameters and hyperparameters used in this work are presented in Table~\ref{tab:lstm_parameters}, unless specified otherwise. The dataset was divided into three parts: training data for model training, validation data to check for possible overfitting and accordingly tune hyperparameters, and test data for performance evaluations.

We begin by examining the convergence behavior of the \gls{dnn} during the training and validation phases to address the risk of significant overfitting. First, we analyze the convergence of trained models with three methods in Fig.~\ref{fig:train_loss}. It shows that the \textbf{RF-Localization} method has a slower convergence compared to \textbf{RF-Blockage} and \textbf{RF+Lidar-Blockage}, potentially due to the high randomness observed in its only input: RSSI data. Note that the loss for \textbf{RF-Localization}  is related to the location of the vehicle, whereas for \textbf{RF-Blockage} and \textbf{RF+Lidar-Blockage}, the loss is associated with the detection of blockages. 

\begin{figure}[t]
    \centering
    \includegraphics[width=0.9\linewidth]{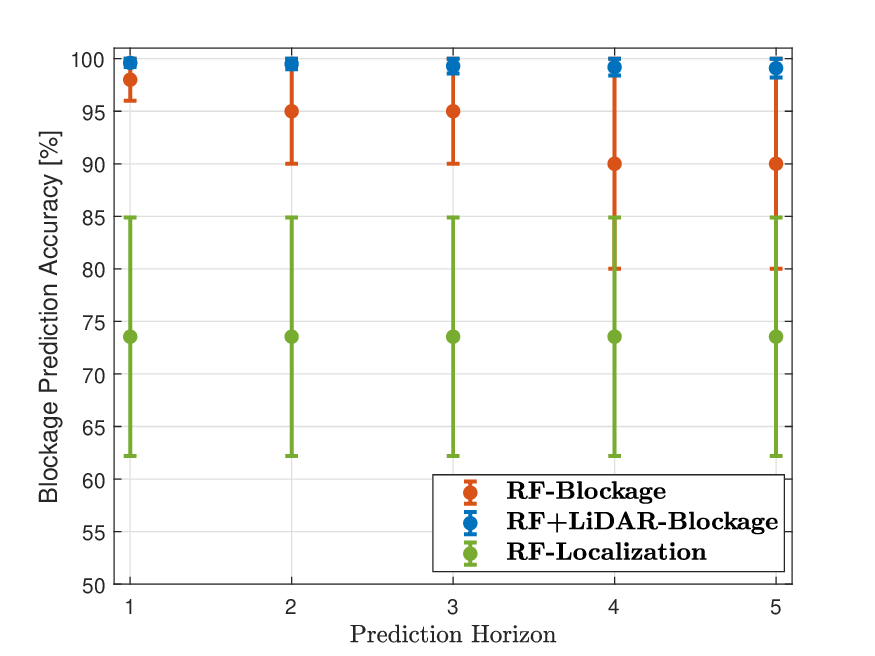}
    \caption{Blockage prediction accuracy using \textbf{RF-Blockage} method, \textbf{RF+LiDAR-Blockage} method and the proposed method, \textbf{RF-Localization}.}
    \label{fig:bar_acc}
\end{figure}

The model's performance is evaluated using the blockage prediction accuracy shown in Fig.~\ref{fig:bar_acc}. The proposed solution achieved an accuracy of approximately $74.47 \%$ and a precision of approximately $86.96 \%$. In comparison, the baseline methods \textbf{RF-Blockage} and \textbf{RF+Lidar-Blockage} have accuracies of approximately $93,60 \%$ and $98.52\%$, respectively. Fig.~\ref{fig:bar_acc} shows blockage prediction accuracy across different prediction horizons, with values generally clustered around $63-85\%$. The error bars indicate variability, with longer horizons displaying greater uncertainty in prediction accuracy, although the central trend remains relatively stable. Note that the variability of the baseline methods increases with the prediction horizon, while the variability of the proposed method remains stable.

\begin{figure}[!t]
    \centering
    \includegraphics[width=0.9\linewidth]{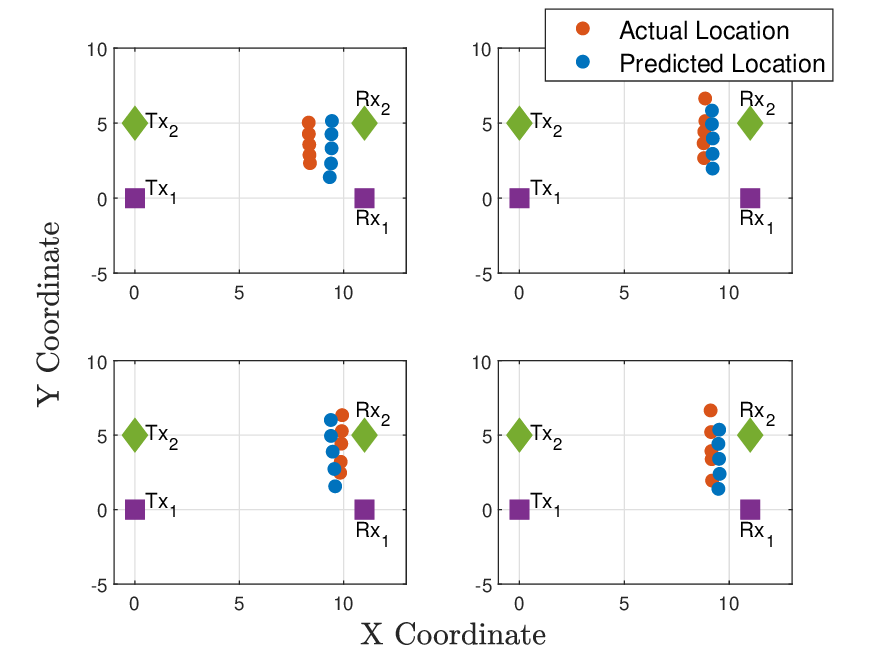}
    \caption{Actual and predicted $5$ subsequent locations of the vehicle centroids for different time windows.}
    \label{fig:pred_actual}
\end{figure}

In Fig.~\ref{fig:pred_actual}, we consider blockage prediction at a newly introduced \gls{tx}-\gls{rx} link without end-to-end retraining. This demonstrates a fundamental limitation of baseline methods like \textbf{RF-Blockage} and \textbf{RF+Lidar-Blockage}, which require extensive retraining to adapt to new \gls{tx}-\gls{rx} configurations. In contrast, our method employs a consistent prediction strategy: it first estimates vehicle centroids' locations and then determines blockage likelihood through geometric analysis. This highlights the generalizability and adaptability of our approach across different \gls{tx}-\gls{rx} pairs, sacrificing a certain degree of prediction accuracy, as observed in Fig.~\ref{fig:bar_acc}.

Fig.~\ref{fig:pred_actual} also compares vehicle centroids' actual and predicted locations over various time windows using the scenario dataset $27$. The close alignment of trajectories illustrates the precision of our model. Furthermore, using the same trained model, we showcase the ability to identify blockage events across multiple \gls{tx}-\gls{rx} pairs. By decoupling location prediction from blockage evaluation, our proposed method provides flexibility, requiring only adjustments in geometric analysis for new configurations, thereby avoiding the costly retraining needed by baseline methods.

\section{Conclusion}\label{sec:conclusion}
In this paper, we introduced a novel approach for predicting blockages in \gls{mmwave} communication systems using self-supervised learning and deep learning on \gls{rf} and \gls{lidar} data. Our method first labels \gls{rf} data with object locations using \gls{lidar} data as input for a \gls{dbscan} algorithm and then applies an \gls{lstm} model to predict location events based solely on the labeled \gls{rf} data. Then, we apply the predicted location in a geometric analysis to verify if the blockages occur. Real-world evaluations using the deepSense6G dataset showed that our model can correctly identify over $74\%$ of blockages, with $86\%$ of positive predictions being accurate. A key strength of this approach is its ability to preemptively identify blockage events and use that information at the transmitter for proactive decision-making, without requiring retraining, in which the same model can also be applied to predict blockages for other \gls{tx}-\gls{rx} links in the surrounding area.
\bibliographystyle{IEEEtran}
\bibliography{mmWaveletter}

\end{document}